\newcommand \exceptfour {E_4}
\newcommand \exceptten {E_{10}}
\newcommand \nsites {L^2}
\newcommand \nbad {M}
\newcommand \nion {V}
\newcommand \spin {S}

\documentstyle[12pt]{article}
\input epsf      

\newcommand \preprint {1}

\newcommand \no {\noindent}
\newcommand \qed{\vrule height5pt width5pt}

\newcommand \bigo {O}

\newcommand \sumxy {\sum_{<x,y>}}
\newcommand \hop {c^\dagger_x c_y }
\newcommand \half {{1 \over 2}}

\newcommand \figgs {1 } 
\newcommand \figfourth {2 } 
\newcommand \figvertices {3 } 

\newtheorem{lemma}{Lemma}
\newtheorem{theorem}{Theorem}

\begin{document}

\title{Phase separation in the neutral Falicov-Kimball model}

\author{Tom Kennedy
\\Department of Mathematics
\\University of Arizona
\\Tucson, AZ 85721
\\ email: tgk@math.arizona.edu
}
\maketitle

\begin{abstract}
The Falicov-Kimball model consists of  spinless electrons
and classical particles (ions) on a lattice. The electrons hop between 
nearest neighbor sites while the ions do not. We consider the model with
equal numbers of ions and electrons and with a large on-site attractive 
force between ions and electrons.
For densities $1/4$ and $1/5$ the ion configuration in the ground state
had been proved to be periodic. 
We prove that for density $2/9$ it is periodic as well.
However, for densities between $1/4$ and $1/5$ other than $2/9$ we prove
that the ion configuration in the ground state is not periodic. 
Instead there is phase separation.
For densities in $(1/5,2/9)$ the ground state ion configuration 
is a mixture of the density $1/5$ and $2/9$ ground state ion configurations.
For the interval $(2/9,1/4)$ it is a mixture
of the density $2/9$ and $1/4$ ground states. 
\end{abstract}

\newpage

\section{Introduction} \setcounter{equation}{0}

The spinless Falicov-Kimball model has two types of particles:
spinless electrons and classical particles, which we refer to as ions.
The particles are on a lattice with the restriction that 
there is at most one ion at each lattice site.
The spinless electrons are fermions, so there is at most one electron
per site.
The electrons can hop between nearest neighbor sites, but the ions cannot.
There is an on-site interaction between electrons and ions.
The Hamiltonian is 
\begin{equation} 
 H = \sumxy \hop - 4 U \sum_x c^\dagger_x c_x \nion_x \label{eqham}
\end{equation}
where $c^\dagger_x$ and $c_x$ are creation and annihilation
operators for the electrons. 
$\nion_x$ is the occupation number for the ions, i.e., $\nion_x= 1$ 
if there is a ion at $x$ and $\nion_x=0$ if there is not. 
The sum over $<x,y>$ is over nearest neighbor bonds in the lattice.
(The factor of 4 in front of the $U$ is included for latter convenience.)
This paper is only concerned with the square lattice, although the
model may be defined on any lattice. 
We will only consider the neutral model in which the number of electrons
is equal to the number of ions, and the interaction parameter $U$ will 
be large and positive. 
By a hole-particle transformation results for positive $U$ imply 
results for negative $U$, but we will not bother to state them.

A review of rigorous work on the Falicov-Kimball model may be found 
in \cite{gm}.
Here we mention only some of the work on the neutral model for large 
positive $U$. 
In one dimension it is expected that for large $U$ the ground state 
of the neutral model with rational density is the periodic arrangement of the 
ions which is ``most homogeneous.''
(There is an explicit algorithm for
determining the most homogeneous configuration.) 
This was proved by Lemberger for $U>U_c$ where $U_c$ depends on the 
denominator of the rational density \cite{l}. 
In any number of dimensions
the ground state for density $1/2$ is the checkerboard configuration
for all $U>0$ \cite{bs,kl}. In two dimensions with large $U$ 
the ground states for densities 
$1/3, 1/4$ and $1/5$ are known rigorously and are periodic \cite{gjlii,k}.
For densities between $1/4$ and $1/2$ there are partial results on the 
ground state \cite{k}, but there is no proof it is periodic for 
rational densities. Based on what is known in one and two dimensions
and the methods used to obtain these results, it is natural to conjecture 
that in two dimensions
the ground state for large $U$ is periodic for rational densities.

In this paper we prove that this conjecture is wrong for densities 
between $1/5$ and $1/4$ other than $2/9$. 
In this density range there is phase separation
in the ground state. The phases involved are the ground states for densities
$1/5$, $2/9$ and $1/4$, which are shown in figure \figgs. The ground 
state for densities between $1/5$ and $2/9$ is made up of large regions of 
density $1/5$ and density $2/9$ ground states 
with the relative areas chosen to yield the desired density. 
A similar statement holds for densities between $2/9$ and $1/4$.
The precise theorem is as follows.

\ifodd \preprint 
{
\epsfbox[0 0 200  240 ] {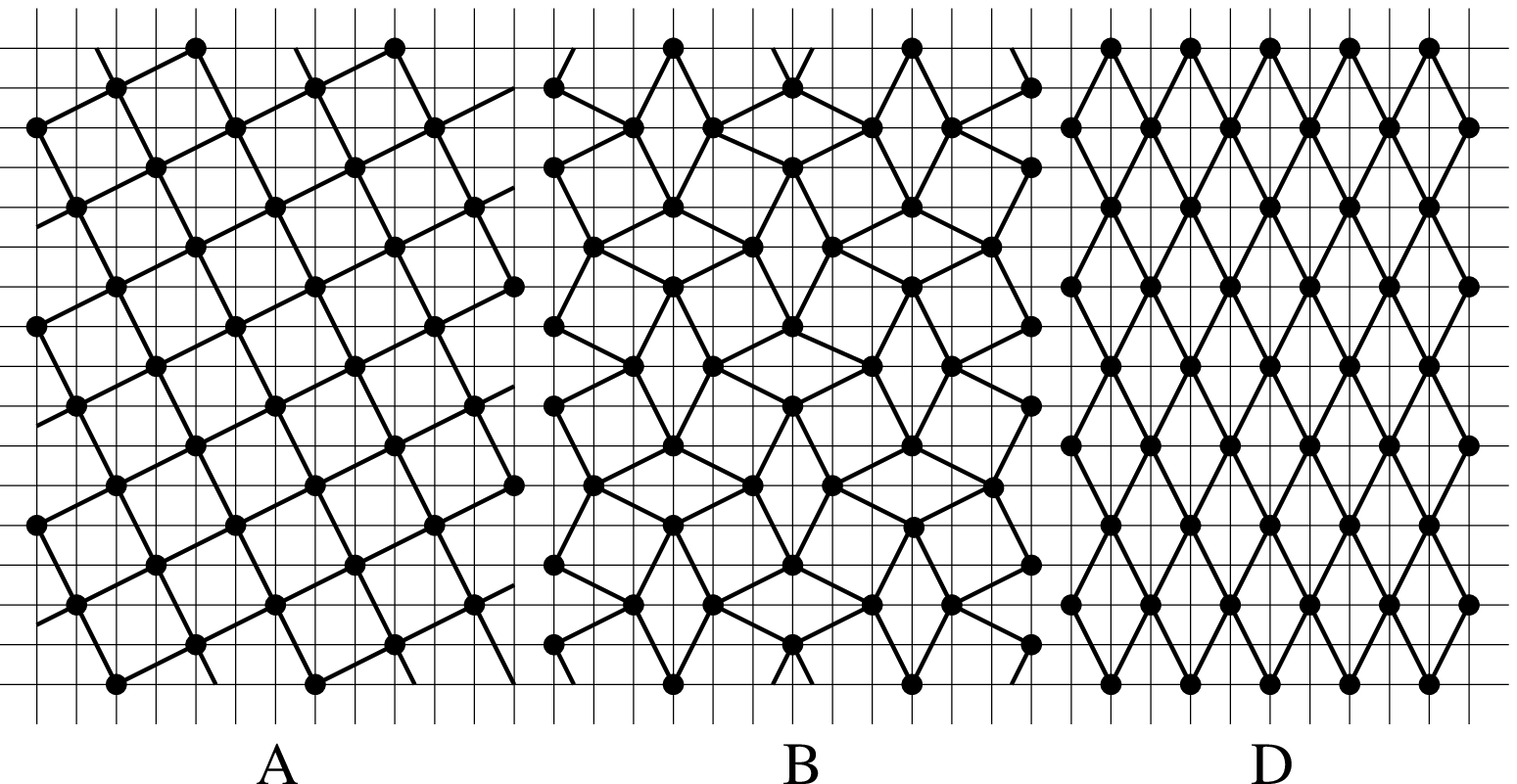} 
{\leftskip=20 pt \rightskip= 20 pt \noindent
\no Figure \figgs : The ground states for densities 1/5, 2/9 and 1/4
(left to right). The heavy lines between the ions are only a guide for
the eye.
\par}

\bigskip
\bigskip
}
\else { } \fi

\begin{theorem} \label {thmmain}
There are positive constants $U_0$ and $c$ 
such that for $U \ge U_0$ the following is true for $L$ 
by $L$ squares $\Lambda$ with $n$ ions and $n$ electrons. 
Let $\rho = n /L^2$ be the common density.
If the density $\rho$ is $2/9$, $L$ is a multiple of 6 and we use periodic
boundary conditions, then the 
ground state configurations of the ions are configuration B in
figure \figgs and its translates.
If the density $\rho$ is between $1/5$ and $2/9$, then for every ground
state configuration of the ions we can find a subset $\Lambda_0$ 
of $\Lambda$ which contains at most $c U^8 L$ sites
and is such that on each connected component of $\Lambda \setminus \Lambda_0$
the configuration agrees with either configuration A or B in 
figure \figgs (up to a lattice symmetry).
If the  density $\rho$ is between $2/9$ and $1/4$, then the same statement
is true with ``A or B'' replaced by ``B or D.'' 
(The statements for densities other than $2/9$ are true for any choice
of boundary conditions.)
\end{theorem}

Although $U^{8}$ is large, the key point is that the bound on the 
number of sites in $\Lambda_0$ contains $L$ while the number of sites
in $\Lambda$ is $L^2$. So for large $L$, $\Lambda_0$ is a tiny 
fraction of the total area. We can think of $\Lambda_0$ as consisting
of domain walls between large regions which contain one of the configurations 
shown in figure $\figgs$. The phase separation 
we find in this range of densities for the neutral model 
should not be confused with the phase separation discussed by 
Freericks and Falicov \cite{ff}. 
Their argument applied to the non-neutral model. 

The expectation that the neutral model with rational density should have 
periodic ground states is based on the following intuition. 
The attraction between electrons and ions is large, so each electron
spends most of its time at a site with an ion. Now consider the 
kinetic energy of an electron. If nearby sites
have ions, then the electrons at those sites will restrict the movement 
of the electron we are considering. So its kinetic energy is 
minimized by spreading out
the ions as much as possible to maximize the space that each electron 
has to move in. However, Watson \cite{w} emphasized that there will typically 
be a mismatch between the lattice and the natural ion configuration 
in the absence of a lattice.
Thus the lattice structure can frustrate the exclusion
principle's attempts to put the ions in the ``most homogeneous'' configuration.
We should emphasize that this paper only covers a small interval of densities.
An important open question is whether the phase separation we find
here holds for most densities, or whether there are intervals in which 
the rational densities have periodic ground states which are the most 
homogeneous in some sense. 

We conclude the introduction by sketching the proof for densities in
$(1/5,2/9)$. 
When $U$ is large and the model is neutral, the ground state energy
of a given ion configuration may be expanded in powers of $1/U$. 
This yields an effective Hamiltonian for the ions. One can begin to study
it by only keeping terms up to a certain order in $1/U$. 
(Of course, to prove anything one must eventually consider all orders.)
Watson \cite{w} showed that when the density is between $1/5$ and $1/4$
the ground states of the fourth order Hamiltonian 
correspond to tilings of the plane by squares and diamonds in 
which the squares and diamonds have the dimensions of those found in 
figure \figgs. 
Watson's result will play a crucial role in our proof. 
The vertices in such a tiling can be one of four types which we label
A, B, C or D following Watson's notation.
The four types are shown in figure \figvertices.
Note that in figure \figgs configurations A, B or D contain only
vertices of type A, B or D, respectively.

All configurations
that correspond to a square-diamond tiling have the same energy through
fourth order. To determine the ground states for densities in $(1/5,2/9)$
we must go to higher orders in the perturbation series. 
Following Watson's treatment of a similar model, we write the 
Hamiltonian as a function of the number of each type of vertex. 
At sixth order these square-diamond tilings still all have the same energy.
At eighth order vertices of type A,B and C have the same energy but
vertices of type D have higher energy. For densities between 
$1/5$ and $2/9$ there are square-diamond tilings which contain no
vertices of type D. To determine the ground state among all these 
configurations we must go to tenth order. Here we find that the energy 
of a type C vertex is higher than that of a type B vertex. Thus for densities
in $(1/5,2/9)$ the ground state must be a square-diamond tiling with only
type A and B vertices. (Note that the tiling with only type A vertices
has density 1/5, while the tiling with only type B vertices has density
2/9. So one can obtain any density in $(1/5,2/9)$ by a suitable 
mix of type A and B vertices.) However, 
a type A vertex cannot be adjacent to a type B vertex. 
Thus we must separate the A and B vertices to minimize the energy.

\vfill \eject

\section{Proof of phase separation} \setcounter{equation}{0}

To derive the perturbation theory it is convenient to change to ``spin''
variables for the ions. 
Let $2\nion_x=\spin_x+1$, so $\spin_x=1$ when there
is an ion at $x$ and $\spin_x= -1$ when there is not. 
We then take 
$$H = \sumxy \hop - 2U \sum_x c^\dagger_x c_x \spin_x$$
This differs from the original Hamiltonian by a term proportional
to $\sum_x c^\dagger_x c_x$, but we will only consider problems in which the 
number of electrons is fixed, so such a term is constant.

There are no interactions between the electrons, so the Hamiltonian
is just the second quantized form of the single electron
Hamiltonian $T - 2US$. The operator $T$ has matrix 
elements $T_{xy}$ with $T_{xy}=1$ if $|x-y|=1$ and $T_{xy}=0$ 
otherwise. The operator $S$ is diagonal with entries $S_x$. 
The ground state energy for $N$ electrons
is the sum of the $N$ lowest eigenvalues of $T-2US$. 
To find the ground state
for a particular density of electrons and ions we must minimize this energy
over all $S$ with the desired ion density.

Let $H(\spin)$ be the ground state energy for the ion
configuration $S$ with the number of electrons equal 
to the number of ions. To expand $H(\spin)$ in powers of $1/U$, we 
begin by rewriting $H(\spin)$ as in \cite{kl}.
If $U>2$ then the number of negative  
eigenvalues of $T-2US$ is equal to the number of sites with $\spin_x= 1$,
i.e., the number of ions.
Thus when the number of electrons equals the number of ions, we
have
\begin{equation}
H(\spin) = \sum_{\lambda_i <0} \lambda_i = \half [Tr(T-2US)-Tr(|T-2US|)] 
\end{equation}
where $\lambda_i$ are the eigenvalues of $T-2US$. 
Now $Tr(T)=0$, and if we keep the number of ions fixed then
$Tr(S)$ is a constant. So we might as well redefine
$H(\spin) = - {1 \over 2}  Tr(|T-2US|)$. 
Then we write this as 
\begin{equation}
H(S) = -{1 \over 2} Tr(|T-2US|) = -{1 \over 2} Tr( [(T-2US)^2]^\half) 
= -U Tr(1+\Delta)^\half
\end{equation}
with
$$\Delta = - (2U)^{-1}(TS+ST) + (2U)^{-2}T^2$$
We have used the fact that $S^2=1$.

Now we derive the perturbation theory by following the treatment
by Gruber, Jedrzejewski and Lemberger \cite{gjlii}. 
A somewhat different derivation may be found in \cite{l}.
If $U$ is sufficiently large, then $||\Delta|| < 1$ and we
may expand $(1+\Delta)^\half$ in a power series in $\Delta$. 
Since $T_{xy}$ is nonzero only if $x$ and $y$ are nearest neighbors, 
when we take the trace of each term we generate nearest neighbor walks
that end where they start. Grouping together terms with the same 
power of $U^{-1}$, we may write the result as 
\begin{equation}
H(\spin)= \sum_{m=1}^\infty \, U^{-2m+1} \, \sum_X \, h_{2m,X} \, \spin_X \label{eqa}
\end{equation}
$X$ is summed over finite subsets of the lattice, and 
$\spin_X = \prod_{x \in X} \spin_x$. The coefficient $h_{2m,X}$ is nonzero only if
there is a nearest neighbor walk with $2m$ steps which ends where it
begins and visits each site in $X$. (It may visit sites outside of $X$ as
well.) The coefficients $h_{2m,X}$ are invariant under the usual 
lattice symmetries. There is a
constant $c$ such that for every site $x$
\begin{equation}
\sum_{X: x \in X} |h_{2m,X}| \le c^m \label{eqb}
\end{equation}

We need to compute this effective Hamiltonian through tenth order, i.e., 
through the $m=5$ terms. There are a lot of terms at this order. Since
we will consider relatively low densities, it will prove useful to go
back to the occupation variables $\nion_x$. 
(Recall that $\spin_x = 2\nion_x-1$.)
Equation (\ref{eqa}) gives
\begin{equation}
 H(\nion)= \sum_{m=1}^\infty \, U^{-2m+1} \, 
  \sum_X \, c_{2m,X} \, \nion_X \label{eqaa}
\end{equation}
where $\nion_X = \prod_{x \in X} \nion_x$.
The coefficients $c_{2m,X}$ may be computed from the $h_{2m,X}$ in a 
straightforward manner. In particular, they are nonzero only if 
$X$ is contained in the set of sites visited by a $2m$ step nearest 
neighbor walk that ends where it begins. The $c_{2m,X}$ satisfy a bound
like (\ref{eqb}). 

For the proof it is useful to split the effective Hamiltonian into three 
parts.
\begin{equation} 
H = H_4 + H_{10} + H_\infty 
\end{equation} 
$H_4$ contains the terms in eq. (\ref{eqaa}) for $m=1$ and $2$. 
$H_{10}$ contains the terms with $m=3,4$ and $5$, and 
$H_\infty$ contains the terms with $m>5$.
The fourth order Hamiltonian $H_4$ has been computed before \cite{gjlii}.
In the occupation variables it may be written in the following form.
\begin{eqnarray*}
H_4 = && U^{-1} [ 8 \sum_{<xy>:|x-y|=1} \nion_x \nion_y - 16 \sum_{x} \nion_x ] \\
 && + U^{-3} [ 64 \sum_{<xy>:|x-y|=\sqrt{2}} \nion_x \nion_y 
 + 16 \sum_{<xy>:|x-y|=2} \nion_x \nion_y - 16 \sum_{x} \nion_x 
 + \sum_{X \in \exceptfour} c_{4,X} \nion_X ] \\
\end{eqnarray*}
$\exceptfour$ is the collections of sets $X$ which appear at fourth order
and contain a pair of sites $x,y$ with $|x-y|=1$. There are only a few
such sets and we can compute their coefficients explicitly, but as we 
will see their actual values play no role in the proof. If $X \in \exceptfour$
and $\nion_X \ne 0$, then there is a nearest neighbor pair $x,y$ with 
$\nion_x \nion_y=1$. This gives a contribution of $8 U^{-1}$ to $H_4$ which is 
much larger than the order $U^{-3}$ contribution from $c_{4,X} \nion_X$. 

It was shown in 
\cite{k} that for configurations with density between $1/5$ and $1/4$ which
minimize $H_4$ every 3 by 3 block of sites must look like one of the 
four configurations shown in figure \figfourth. 
(We will include a proof of this later.)
Watson \cite{w} made the following observation that will play an essential 
role in our proof.

\ifodd \preprint 
{
\epsfbox[0 0 200  120 ] {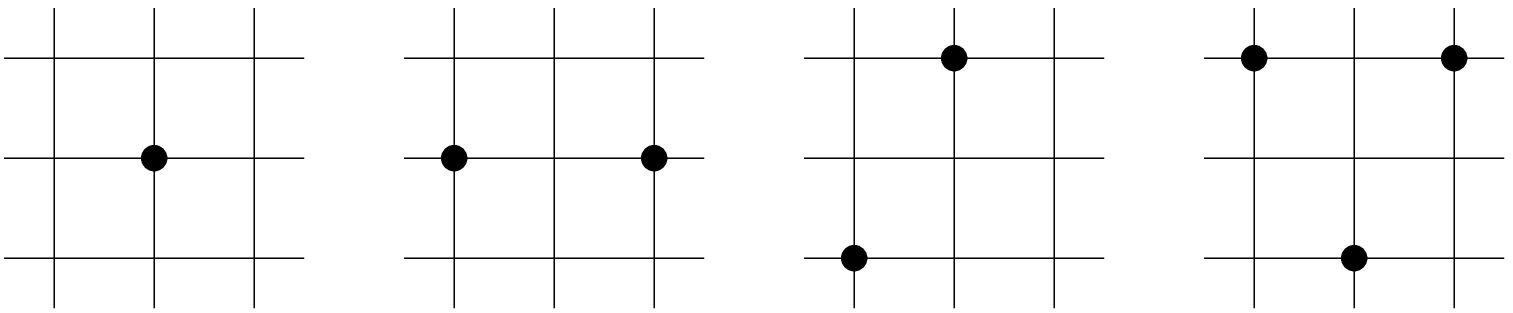} 

{\leftskip=20 pt \rightskip= 20 pt \noindent
\no Figure \figfourth : For densities between 1/5 and 1/4, 
fourth order perturbation theory implies
that in a ground state every 3 by 3 square must be one
of the above cases, up to lattice symmetries. 
\par}

\bigskip
\bigskip
}
\else { } \fi

\begin{lemma} \label {lema} 
If every 3 by 3 block equals one of those shown in figure \figfourth
(up to a lattice symmetry), then the configuration corresponds to 
a tiling of the lattice by squares and diamonds in which every vertex looks
like one of the four types shown in figure \figvertices up to a lattice 
symmetry.
\end{lemma}

\ifodd \preprint 
{
\epsfbox[0 0 200  150 ] {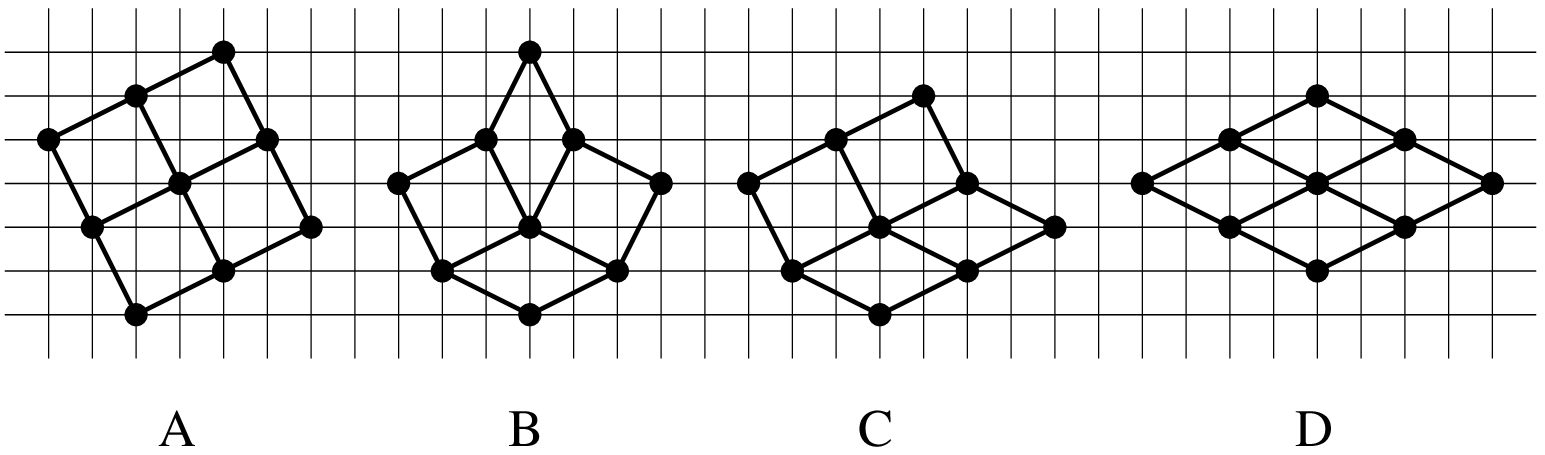} 
{\leftskip=20 pt \rightskip= 20 pt \noindent
\no Figure \figvertices : For densities between 1/5 and 1/4, every 
ground state corresponds to a tiling of the plane by squares and diamonds
in which every vertex must be one of the four types shown above. 
The ions at the center of each of these figures are called type 
A, B, C or D ions, respectively.
\par}

\bigskip
\bigskip
}
\else { } \fi

\no {\bf Proof \cite{w}:} Note that the 3 by 3 blocks we are using can
overlap. Thus the condition that every one of these blocks is one of 
those shown in figure \figfourth puts many constraints on the configuration.
We start with a site at which there is an ion.
Using the fact that every 3 by 3 block is one of those shown in figure
\figfourth we work outwards from the initial ion and determine all possible
configurations in the neighborhood of the inital ion. We find 
that the configuration must look like one of the four cases
shown in figure \figvertices. \qed

\medskip

It is important to note that the proof of the lemma is local. If we have
a configuration in which some 3 by 3 blocks are not one of those 
shown in figure \figfourth, then we can still conclude that 
in the parts of the lattice where the 3 by 3 
blocks are one of those shown in the figure the configuration must be
given by a square-diamond tiling.
We will say that an ion is type A, B, C or D if the configuration
in the neighborhood of the ion agrees with that shown in figure \figvertices
up to a lattice symmetry.
Note that in regions where the configuration does not correspond to 
a square-diamond tiling there will be ions that are not any of these
four types.

The tenth order Hamiltonian $H_{10}$ contains too many terms to list here.
However, we can organize the Hamiltonian so that we only need the actual 
values of the coefficients of a modest number of them. 
The values of the coefficients of the terms in $H_{10}$ are important only
in regions where $H_4$ is minimized. By the lemma the configuration 
corresponds to a square-diamond tiling in such regions. In these tilings
two ions cannot be separated by a distance $1,\sqrt{2}$, $\sqrt{8}$ or $3$. 
So we define $\exceptten$ to be
the collection of sets that appear at tenth order or lower and contain 
a pair of sites $x,y$ with $|x-y|^2$ equal to $1,2,8$ or $9$. 
For $X \notin \exceptten$ the coefficients $c_{6,X},c_{8,X},c_{10,X}$
are given in table 1. 

\ifodd \preprint 
{
\bigskip
\bigskip

\begin{tabular}{||l||r|r|r||r|r|r|r||} \hline
 term                       & 6th & 8th   & 10th   & A &  B  &  C  & D \\
 \hline
 (0,0)                      &  64 &   112 &   -704 & 1 &  1  &  1  & 1 \\ 
 (0,0), (2,0)               &  96 & -1360 &  -1440 & 0 & 1/2 & 1/2 & 1 \\ 
 (0,0), (2,1)               & 216 &  -768 & -14000 & 2 &  2  &  2  & 2 \\
 (0,0), (3,1)               &     &   512 &    640 & 2 &  1  &  1  & 0 \\ 
 (0,0), (4,0)               &     &    32 &   1120 & 0 & 1/2 & 1/2 & 2 \\ 
 (0,0), (3,2)               &     &       &   4000 & 0 &  0  & 1/2 & 2 \\ 
 (0,0), (4,1)               &     &       &   1000 & 0 &  2  &  1  & 0 \\ 
 (0,0), (5,0)               &     &       &     40 & 2 &  0  & 1/2 & 0 \\ 
 (0,0), (2,0), (1,2)        &     & -2592 &  10560 & 0 &  1  &  1  & 2 \\ 
 (0,0), (2,0), (4,0)        &     &   192 &   1760 & 0 &  0  &  0  & 1 \\ 
 (0,0), (2,0), (1,3)        &     &       &  -7040 & 0 &  0  &  0  & 0 \\ 
 (0,0), (2,0), (3,2)        &     &       &  -5280 & 0 &  0  &  1  & 4 \\ 
 (0,0), (2,1), (4,0)        &     &       &  -3960 & 0 &  1  &  1  & 2 \\ 
 (0,0), (2,1), (4,1)        &     &       &   1320 & 0 &  2  &  1  & 0 \\ 
 (0,0), (2,1), (-1,2)       &     &       & -15840 & 4 &  2  &  2  & 0 \\ 
 (0,0), (1,2), (2,0), (3,2) &     &       &  15840 & 0 &  0  & 1/2 & 2 \\ 
\hline 
\end{tabular}

\bigskip

{\leftskip=20 pt \rightskip= 20 pt \noindent
\no Table 1 : The sixth, eighth and tenth order terms in the 
Hamiltonian when restricted to 
configurations that correspond to tilings by squares and diamonds.
\par}

\bigskip
}
\else { } \fi

Now we are ready to state and prove the main inequalities that will
show there is phase separation for densities between $1/5$ and $1/4$ 
other than $2/9$. 

\ifodd \preprint 
{
\vfill
\eject
}
\else { } \fi

\begin{theorem} \label {thma}
Let $\nbad$ be the number of 3 by 3 blocks in which the configuration is
not one of those shown in figure \figfourth. Let $n_A, n_B, n_C$
and $n_D$ be the number of type A, B, C and D ions respectively 
(figure \figvertices). 
Consider an $L$ by $L$ lattice so there are 
$\nsites$ sites, and let $\rho$ be the ion density so there are 
$\rho \nsites$ ions.
Then there are polynomials $p_1,p_2,\cdots,p_6$ in $U^{-1}$ and
functions $f_1,f_2,f_3,f_4$ of $U^{-1}$ such that 
\begin{equation}
 {16 \over 3} U^{-3} \nbad  \le 
H_4 - p_1 \rho \nsites - p_2 \nsites 
= \bigo(U^{-1}) \nbad
\label{inequalfour}
\end{equation}
\begin{equation}
H_{10} - p_3 \rho \nsites - p_4 \nsites
 = 224 U^{-7} n_D + 2340 U^{-9} n_C + 12240 U^{-9} n_D + \bigo(U^{-5}) \nbad
\label{inequaltena}
\end{equation}
\begin{equation}
H_{10} - p_5 \rho \nsites - p_6 \nsites
 = 224 U^{-7} n_A + 2340 U^{-9} n_C + 12240 U^{-9} n_A + \bigo(U^{-5}) \nbad
\label{inequaltenb}
\end{equation}
\begin{equation}
H_\infty - f_1 \rho \nsites - f_2 \nsites = \bigo(U^{-11}) (n_C + n_D + \nbad)
\label{inequalinfinitya}
\end{equation}
\begin{equation}
H_\infty - f_3 \rho \nsites - f_4 \nsites = \bigo(U^{-11}) (n_A + n_C + \nbad)
\label{inequalinfinityb}
\end{equation}
$\bigo(U^{-k})$ denotes a quantity whose absolute value may be bounded by 
a constant times $U^{-k}$. 
\end{theorem}

\no {\bf Proof:} The inequality for $H_4$ was proved in \cite{k}. 
We include a short proof for completeness. Let $B$ be a 3 by 3 block of 
sites. (So it contains 9 sites.) Let $z$ be the center of $B$. Define
\begin{eqnarray*}
H_B && = U^{-1} {4 \over 3} \sum_{<xy> \subset B: |x-y|=1} \nion_x \nion_y
 + U^{-3} [16 \sum_{<xy> \subset B: |x-y|=\sqrt{2}} \nion_x \nion_y
 + {16 \over 3} \sum_{<xy> \subset B: |x-y|=2} \nion_x \nion_y \\
 && - 16 \nion_z
 - {32 \over 3} \sum_{x \in B:|x-z|=1} \nion_x 
 - {16 \over 3} \sum_{x \in B:|x-z|=\sqrt{2}} \nion_x +16
 + \sum_{X \subset B, X \in \exceptfour} \, {c_{4,X} \, \nion_X \over m_X} ]
 \\
\end{eqnarray*}
where $m_X$ is the number of translates of $X$ that are contained in $B$. 
Now consider $\sum_B H_B$ where the sum is over all 3 by 3 blocks. 
(So some of them overlap.) A pair $<xy>$ with $|x-y|=1$ is contained in 
6 different blocks, a pair with $|x-y|=\sqrt{2}$ in 4 different blocks, 
and a pair with $|x-y|=2$ in 3 different blocks. 
Using these facts we find 
\begin{equation}
 \sum_B H_B = H_4 + (16 U^{-1} - 64 U^{-3}) \rho \nsites + 16 U^{-3} \nsites
\end{equation}
So to complete the proof we must show that $H_B$ vanishes if $B$ is one
of the blocks shown in figure \figfourth and is at least ${16 \over 3} U^{-3}$
otherwise. First note that if the block contains a pair of nearest neighbor
ions then the second order part of $H_4$ is at least 
${4 \over 3} U^{-1}$. Since the fourth order part is smaller by a factor 
of $U^{-2}$, this shows that $H_B \ge {16 \over 3} U^{-3}$ for all such 
configurations if $U$ is large enough. Now suppose that the block 
does not contain any nearest neighbor ions. Note that this implies that
$\nion_X=0$ for all $X \in \exceptfour$. So we can easily compute $H_B$ for
such configurations. We find that 
for all 3 by 3 blocks that do not contain a nearest neighbor
pair of ions and which are not in figure \figfourth, 
$H_B$ is at least ${16 \over 3} U^{-3}$. $H_B$ vanishes on 
the configurations in  figure \figfourth.
This completes the proof of (\ref{inequalfour}).

To prove (\ref{inequaltena}) and (\ref{inequaltenb}) 
we first consider configurations which 
correspond to a square-diamond tiling. If $X \in \exceptten$ then
$\nion_X=0$ in these configurations. We will consider two $X$'s to be 
equivalent if they are related by a translation, reflection and or 
rotation. For $X \notin \exceptten$ there are 16 equivalence classes, 
listed in table 1. For each equivalence class we want to write the
number of $X$ in the class with $\nion_X = 1$ in terms of 
$n_A, n_B, n_C, n_D$. 
Consider the second equivalence class. It contains those $X$'s
of the form $X=\{x,y\}$ with $|x-y|=2$. The number of such $X$ with 
$\nion_X \ne 0$ in figure \figvertices
is 0,2,2, or 4 for A,B,C, or D, respectively. However, this 
overcounts the number of $X$ with $\nion_X \ne 0$. Each $X$ is counted 
4 times. So the number of $X$'s in the second equivalence class with 
$\nion_X = 1$ is $0 n_A + {1 \over 2} n_B + {1 \over 2} n_C + n_D$.
These coefficients  $0,{1 \over 2},{1 \over 2},1$ along with 
the coefficients for all the other equivalence classes are given 
in table 1. We should emphasize that the overcounting factor is not
always 4. It varies from equivalence class to equivalence class,
and in one case within the equivalence class. 
Using table 1 we find that for configurations which correspond to 
a square-diamond tiling,
\begin{eqnarray*}
H_{10} &&= U^{-5} ( 496   n_A   +544 n_B   +544 n_C  +592 n_D) \\
      && + U^{-7} (-400   n_A  -4168 n_B  -4168 n_C -7712 n_D) \\
      && + U^{-9} (-90704 n_A -48664 n_B -46324 n_C +5616 n_D) \\
\end{eqnarray*}
The quantities $n_A, n_B, n_C$ and $n_D$ are not all independent. 
Each ion corresponds to a vertex in the square-diamond tiling, so
\begin{equation}
n_A + n_B + n_C + n_D = \rho \nsites \label{abcdn}
\end{equation}
By considering the areas associated with each of the four types of vertices
in figure \figvertices, we see that 
\begin{equation}
5 n_A + {9 \over 2} n_B + {9 \over 2} n_C + 4 n_D = \nsites \label{abcdv}
\end{equation}
Using these two equations we can eliminate $n_A$ and $n_B$ from our
expression for $H_{10}$. 
The result is (\ref{inequaltena}). If we use the two equations to eliminate
$n_B$ and $n_D$, then the result is (\ref{inequaltenb}).

This proves (\ref{inequaltena}) and (\ref{inequaltenb}) 
for configurations that correspond to 
a square-diamond tiling. Now consider a configuration that does not.
The number of sites in the region where it does not correspond to such a 
tiling is at most $9\nbad$. Thus the terms in $H_{10}$ that 
intersect this region can contribute at most $\bigo(U^{-5}) \nbad$. 
Outside of this region we can apply the above argument. There will 
be errors at the boundary of the region in which there is a 
square-diamond tiling since there will be incomplete squares and diamonds,
but the contribution of these errors is also $\bigo(U^{-5}) \nbad$. 
Equations (\ref{abcdn}) and (\ref{abcdv}) are not true for a general 
configuration, but the difference between the right and left side of 
these equations is bounded by a constant times $\nbad$. 
This completes the proof of (\ref{inequaltena})
and (\ref{inequaltenb}).

Finally, we must prove (\ref{inequalinfinitya}) and (\ref{inequalinfinityb}). 
We only give the proof of (\ref{inequalinfinitya}). A similar argument
proves (\ref{inequalinfinityb}), or it may be obtained from  
(\ref{inequalinfinitya}) by showing $|n_A - n_D - 2 \nsites + 9 \rho \nsites|$ 
is bounded by a constant times $\nbad$. 
To compare a term $c_{2m,X} \nion_X$ in $H_\infty$ with the corresponding 
term for the configurations in figure \figgs, we need to know if $X$ is
contained in a region where the configuration agrees with one of the 
configurations in figure \figgs. So we make the following definitions.
For a lattice site $x$ and positive integer $r$, let $B_r(x)$ be the 
set of sites such that the $l^1$ distance from $x$ to $y$ is at most $r$. 
Given a configuration, let $A_m$ be the set of sites with a type A ion
such that $B_m(x)$ is a subset of configuration A in figure \figgs.
$B_m$ is defined analogously using configuration B.
Define $\hat h_{2m,X}= h_{2m,X}/|X|$ so that
\begin{equation}
H_\infty = \sum_{m=6}^\infty U^{-2m+1} \, \sum_x \sum_{X: x \in X} \hat h_{2m,X} \nion_X 
\end{equation}
Since $\nion_X=0$ if $\nion_x=0$, we can restrict the sum over $x$ to sites 
with an ion. 

Let $I$ denote the sites with an ion. We start by comparing $H_\infty$ with 
the sum over a subset of the terms in $H_\infty$:
\begin{eqnarray*}
|H_\infty-\sum_{m=6}^\infty U^{-2m+1} \, \sum_{x \in A_m \cup B_m}
\, \sum_{X: x \in X} \hat h_{2m,X} \nion_X | 
 && = |\sum_{m=6}^\infty U^{-2m+1} \, \sum_{x \in I \setminus (A_m \cup B_m)}
  \, \sum_{X: x \in X} \hat h_{2m,X} \nion_X | \\
\end{eqnarray*}
\begin{equation}
 \le \sum_{m=6}^\infty U^{-2m+1} \, |I \setminus (A_m \cup B_m)| c^m 
\label{hamdif}
\end{equation}
using (\ref{eqb}). We need to estimate the size of 
$I \setminus (A_m \cup B_m)$. Let $x \in I \setminus (A_m \cup B_m)$.
Then $B_m(x)$ is not a subset of configuration A or B in figure \figgs.
So either (i) $B_m(x)$ is not a subset of any square-diamond tiling,
(ii) $B_m(x)$ is a subset of a square-diamond tiling which contains a type 
C or type D vertex, or (iii) $B_m(x)$ is a subset of a square-diamond 
tiling which contains both type A and type B vertices. 
In case (i) $B_m(x)$ must
intersect a 3 by 3 block that is not one of those in figure \figfourth. 
Note that it is not possible for
a tiling to contain both type A and type B vertices, but no type C or D, so
case (iii) never happens.  
Thus we can associate with each site in $I \setminus (A_m \cup B_m)$
either a ``bad'' 3 by 3 block or a type C or type D vertex. 
The number of sites that are associated with the same 3 by 3 block 
or type C or type D vertex is bounded by $d m^2$ for some constant $d$.  
Thus
\begin{equation}
|I \setminus (A_m \cup B_m)| \le d m^2 (n_C + n_D + \nbad) \label{setbound}
\end{equation}
Thus (\ref{hamdif}) is 
\begin{equation}
\le \sum_{m=6}^\infty U^{-2m+1} c^m d m^2 (n_C + n_D + \nbad) = \bigo(U^{-11}) 
 (n_C + n_D + \nbad)
\end{equation}

Let $e_A$ be the energy per site from $H_\infty$ when the entire finite 
lattice contains configuration A in figure \figgs,
and let $e_B$ be the same quantity for configuration B in figure \figgs. 
Let 
\begin{equation}
5 e_A = \sum_{m=6}^\infty U^{-2m+1} a_{2m} 
\end{equation}
If $x \in A_m$ then the definition of $A_m$ implies that 
\begin{equation}
\sum_{X : x \in X}  \hat h_{2m,X} \nion_X = a_{2m}
\end{equation}
Thus 
\begin{equation}
|\sum_{m=6}^\infty U^{-2m+1} \, \sum_{x \in A_m} \sum_{X: x \in X} \hat h_{2m,X} \nion_X 
 - 5 n_A e_A | 
\le \sum_{m=6}^\infty U^{-2m+1} \, \sum_{x \in A \setminus A_m} | a_{2m} |
\label{adif}
\end{equation}
Since $A \setminus A_m \subset I \setminus (A_m \cup B_M)$, (\ref{setbound})
implies $|A \setminus A_m| \le c m^2 (n_C + n_D + \nbad)$. This 
shows that (\ref{adif}) is $\bigo(U^{-11}) (n_C + n_D + \nbad)$.

A similar argument shows 
\begin{equation}
|\sum_{m=6}^\infty U^{-2m+1} \, \sum_{x \in B_m} \, \sum_{X : x \in X}  \hat h_{2m,X} \nion_X 
- {9 \over 2} e_B n_B  |
= \bigo(U^{-11}) (n_C + n_D + \nbad) \label{bdif} 
\end{equation}
Combining our bounds on (\ref{hamdif}) and (\ref{adif}) with (\ref{bdif}) shows
\begin{equation}
|H_\infty- 5 e_A n_A - {9 \over 2} e_B n_B |
 = \bigo(U^{-11}) (n_C + n_D + \nbad)
\end{equation}
which proves (\ref{inequalinfinitya}). 
\qed

The main theorem follows easily from the inequalities we have proved 
by a ``variational'' argument. 

\medskip

\no {\bf Proof of theorem \ref{thmmain}:} 
First we prove the statement about the ground state for density $2/9$.
Configuration B has $M=0$ and only type B vertices, so theorem \ref{thma}
immediately implies it is a ground state. 
Any other ground state must have $M=0$ and no type C or D vertices. 
The condition $M=0$ implies the configuration is a square diamond
tiling. The only tiling which contains only A and B vertices and has
density $2/9$ is configuration B in figure \figgs. 

Now we turn to the proof of the statements for densities between 
$1/4$ and $1/5$. 
We will construct a trial configuration with relatively low energy and
then use the inequalities in theorem \ref{thma} to prove 
theorem \ref{thmmain}. We give the proof 
for the case of densities in $(1/5,2/9)$. The proof for densities
in $(2/9,1/4)$ is similar. 

Divide the $L$ by $L$ square into two rectangles and put configuration A 
in figure \figgs on one side and configuration B on the
other side. The relative areas of the two rectangles are chosen to give
the desired density. The number of 3 by 3 blocks which do not agree 
with one of those in figure \figfourth is bounded by a constant times $L$. 
Thus $H_4 - p_1 \rho \nsites - p_2 \nsites$ is $\bigo(U^{-1}) L$. 
$n_C$ and $n_D$ are both zero, so 
$H_{10} - p_3 \rho \nsites - p_4 \nsites$ 
is $\bigo(U^{-5}) L$ and 
$H_\infty - f_1 \rho \nsites - f_2 \nsites$ 
is $\bigo(U^{-11}) L$.
Any ground state must have energy no greater than that of our trial 
configuration, so theorem \ref{thma} implies that in a ground state 
\begin{eqnarray*}
&& {16 \over 3} U^{-3} \nbad  + 224 U^{-7} n_D + 2340 U^{-9} n_C 
+ 12240 U^{-9} n_D + \bigo(U^{-5}) \nbad \\
&& + \bigo(U^{-11}) (n_C + n_D + \nbad) \le \bigo(U^{-1}) L \\
\end{eqnarray*}
which may be rearranged as 
\begin{eqnarray*}
&& [ {16 \over 3} U^{-3} + \bigo(U^{-5}) + \bigo(U^{-11}) ] \nbad
 + [ 224 U^{-7} + 12240 U^{-9} + \bigo(U^{-11}) ] n_D  \\
&& + [ 2340 U^{-9} + \bigo(U^{-11}) ] n_C
\le \bigo(U^{-1}) L \\
\end{eqnarray*}
If $U$ is large enough this implies that $\nbad + n_C + n_D$ is bounded by 
a constant times $U^8 L$. Now take $\Lambda_0$ to be the union of all
the 3 by 3 blocks which do not agree with one of the configurations 
in figure \figfourth together with the union over all type C and 
type D ions of the region associated with the ion shown in figure 
\figvertices. In $\Lambda \setminus \Lambda_0$ the configuration
must be a square diamond tiling with no C or D vertices. 
Note that type A and type B vertices cannot be adjacent in the tiling,
i.e., separated by a distance $\sqrt{5}$. 
However, this does
not quite insure that each component contains only type A or type B
vertices. For example one can have a component that consists of two 
large regions which are connected only by a single line of sites. 
Then one can have type A vertices on one side of the narrow connection 
and type B on the other side. To eliminate this, redefine $\Lambda_0$ 
to be the original $\Lambda_0$ plus all sites within a distance $d$ 
of the original set. If $d$ is chosen large enough we eliminate the
above problem and every connected component of $\Lambda \setminus \Lambda_0$
will be a square-diamond tiling with only type A or type B vertices. 
\qed

\bigskip
\bigskip

\no {\bf Acknowledgements: } Ideas from \cite{w} play a crucial 
role in this paper. The author thanks Prof. Watson for sending him 
a copy of this paper prior to its publication.
This work was supported in part by NSF grant DMS-9623509.


\begin{thebibliography}{99}

\newcommand \jtype{\it}

\def \jsp    {{\jtype J. Stat. Phys.}\ }
\def \pr     {{\jtype Phys. Rev.}\ }
\def \prb    {{\jtype Phys. Rev. B}\ }
\def \prl    {{\jtype Phys. Rev. Lett.}\ }
\def \cmp    {{\jtype Commun. Math. Phys.} \ }
\def \jpc    {{\jtype J. Phys.: Condens. Matter}\ }
\def \jpa    {{\jtype J. Phys. A: Math. Gen.}\ }
\def \jpcold {{\jtype J. Phys. C: Solid State Phys.}\ }
\def \pl     {{\jtype Phys. Lett.}\ }
\def \lmp    {{\jtype Lett. Math. Phys.}\ }
\def \npb    {{\jtype Nucl. Phys. B}\ }
\def \jmp    {{\jtype J. Math. Phys.}\ }
\def \jap    {{\jtype J. Appl. Phys.}\ }
\def \jpsj   {{\jtype J. Phys. Soc. Jpn.}\ }
\def \rmp    {{\jtype Rev. Math. Phys.}\ }
\def \epl    {{\jtype Europhys. Lett.}\ }
\def \aihp   {{\jtype Ann. Inst. H. Poincar\'e}\ }
\def \jfa    {{\jtype J. Funct. Anal.}\ }
\def \zpcm   {{\jtype Z. Phys. B}\ }
\def \sc     {{\jtype Science}\ }

\newcommand \arctx{archived in Texas mp\_arc }
\newcommand \arclanl{archived in xxx.lanl }

\bibitem{bs} U. Brandt, R. Schmidt, Ground state properties of a 
spinless Falicov-Kimball model.
\zpcm {\bf 67}, 43 (1986).

\medskip

\bibitem{ff} J. K. Freericks, L. M. Falicov, Two-state one-dimensional
spinless Fermi gas. \prb {\bf 41}, 2163 (1990).

\medskip

\bibitem{gjlii} C. Gruber, J. Jedrzejewski, P. Lemberger, 
Ground states of the spinless Falicov-Kimball model. II
\jsp {\bf 66}, 913 (1992).

\medskip

\bibitem{gm} C. Gruber, N. Macris, The Falicov-Kimball model: a review of
exact results and extensions, preprint (1996).

\medskip

\bibitem{k} T. Kennedy,
Some rigorous results on the ground states of the Falicov-Kimball model. 
{\jtype Rev. Math. Phys.} {\bf 6} 901-925 (1994).
Also in {\it The State of Matter}, Michael Aizenman and Huzihiro Araki
(eds.) World Scientific, 1994.

\medskip

\bibitem{kl} T. Kennedy, E. H. Lieb, 
An itinerant electron model with crystalline or
magnetic long range order. {\jtype Physica } {\bf 138A}, 320-358 (1986).

\medskip

\bibitem{l} P. Lemberger, Segregation in the Falicov-Kimball Model.
\jpa {\bf 25}, 715 (1992).

\medskip

\bibitem{w} G. I. Watson, Repulsive particles on a two-dimensional lattice.
Preprint (1996).

\end{thebibliography}
\end{document}